\documentclass{elsart}
\usepackage{psfig}

\begin{document}
\begin{frontmatter}
\title{Adatoms and nanoengineering of carbon}

\author[sussex]{C. P. Ewels}
\author[sussex]{M. I. Heggie}
\author[newcastle]{P. R. Briddon}

\address[sussex]{CPES, University of Sussex, Falmer, Brighton, BN1 9QJ, UK}

\address[newcastle]{Department of Physics, University of Newcastle, Newcastle,
         NE1 7RU, UK}

\begin{abstract}

We present a new and general mechanism for inter-conversion of carbon
structures via a catalytic exchange process, which operates under
conditions of Frenkel pair generation.  The mechanism typically lowers
reaction barriers by a factor of four compared to equivilent 
uncatalysed reactions.  We examine the relevance of this mechanism for
fullerene growth, carbon onions and nanotubes, and dislocations in irradiated
graphite.
\\
Corresponding author. e-mail: c.p.ewels@sussex.ac.uk fax: +44-1273-677196

\end{abstract}

\end{frontmatter}


Nanotechnology with carbon is made possible by the myriad metastable
structures afforded by three fold coordinated carbon. The
quintessential fullerene structure: hexagonal rings with 12 pentagons,
comprises but a small subset of possible metastable structures of
different ring statistics and properties\cite{AoF}.  Any arbitrary shape
of surface may be engineered from a hexagonal net modified to include
rings of different sizes \cite{kroto}.
Quasi-spherical carbon onions\cite{iijima_80,ugarte_92} are observed
after $>$ 100 keV electron irradiation of polyhedral carbon
particles\cite{Zwanger1996} and under some CVD
conditions\cite{Arnault}. They shrink under sustained
irradiation\cite{banhart_99}, suffer self-compression and, through surface
tension, adopt nearly spherical shape. The phenomenon arises from
Frenkel pair creation, migration and
annihilation\cite{sigle_97} and it appears to be a radial
analogue of the {\bf c} axis expansion (or ``plating out''),
observed in graphites under neutron irradiation \cite{kelly}.
One of the structural models which appears
to mimic well experimental HRTEM onion images,
achieves spherical shape through insertion of pentagon/heptagon pairs
or pentagon-pentagon-octagon triads into the ground state,
icosahedral, faceted fullerene shell\cite{maiti}. 

In order to discuss shape changes, it is instructive to map a flat 
graphite sheet to a general shape by the introduction of topological
defects. Pentagons and heptagons are, respectively, positive and 
negative 60$^o$ wedge disclinations. Their disposition is the main
determinant of shape. They can be moved by the absorption and emission
of edge dislocations, which themselves are identified as wedge disclination
dipoles and appear as pentagon-heptagon pairs.

If such processes occur in even-numbered systems through the Stone-Wales (SW)
route\cite{SW}, high barriers must be overcome ($>$ 6
eV)\cite{science}. This may be theoretically possible within the thermal
spike of an irradiation event or under the extreme conditions of a
carbon arc, but is unlikely to contribute as strongly as a mechanism
which operates more homogeneously and is modestly thermally activated
at typical substrate and annealing temperatures.

Carbon dimers afford an intuitively simple basis for interpretation of
growth and attrition of fullerenes (observed as stable, even $n$
clusters C$_n$).  Addition of a C$_2$ dimer to a fullerene or a
graphite sheet followed by bond reconstruction (as invoked in the
``Pentagon Road'' of fullerene growth) leads to replacement of a
hexagon by a pair of pentagons. Removal of such a pair (as might be
invoked in onion formation) yields, in the case of graphite, two
pentagons and an octagon. Each such addition/removal step must be
accompanied by rearrangement to lower energy and stabilise the growth
or shrinkage process.

Carbon dimers are known to be ejected during thermal decomposition of
fullerenes\cite{earlymassspec} and C$_2$ is a substantial component of
a carbon gas\cite{carbongas}. However, odd numbered species
principally C and C$_3$ also exist in the gas phase and may be ejected
under irradiation above the displacement threshold.

These observations led to the proposal that the SW transformations
occurring during growth be auto-catalysed --- catalysed by the
presence of additional carbon atoms {\it i.e.} in locally odd-numbered
systems. Here we explore further this possibility, first discussed by
Eggen {\it et al.}\cite{science} in the case of the last in the
sequence of transformations \cite{austin} which convert an arbitrary
C$_{60}$ fullerene into icosahedral C$_{60}$ or buckminsterfullerene
(BF). In that first study, a carbon adatom was shown to reduce the
barrier for this last transformation step (from a C$_{2v}$ isomer to
BF) from 4.7 to 2.9 eV. Whereas there has been much work on the
topology and sequence of transformations, including a detailed study
of the energy landscape\cite{WalshWales}, there is still a need to
establish energetically viable transformation paths of the kind we
propose here.

As before, we apply Local Spin Density Functional Theory code AIMPRO
\cite{general_aimpro}, using the Ceperley-Alder
functional\cite{ceperley} and norm-conserving
pseudo-potentials\cite{bhs}. A real-space $sp$ basis of 16 Gaussians
per atom is used for the wavefunction and 4 Gaussians per atom for the
valence charge density. Analytic atomic forces are used within the
conjugate gradient method to optimise structures until forces were
negligible.

In this way we have examined the final annealing transformation in
C$_{60}$ when the adatom is not in the lowest energy state and we
have considered generalised transformations that create and move
dislocations and disclinations ({\it i.e.} pentagons and heptagons) in
graphene sheets. The result is a general and low energy route for
nano-engineering carbon structures.


The SW transformation in BF involves the bond linking two pentagons
rotating through 90$^o$ about its centre, effectively causing pairs of
pentagons and hexagons to swap sites (Fig~1a). The new
auto-catalytic exchange mechanism we find (Fig~1b) achieves the
same end, while at the same time the adatom is incorporated into the
ball and a host atom is ejected.  The route and its activation barrier
were obtained by optimisation, removing degrees of freedom
corresponding to atomic movements which simultaneously break one bond
while making another. For the catalysed exchange process the degree of
freedom removed is the difference in squares of AB and AC bond
lengths, {\it i.e.} $r_{A-B}^2$ - $r_{A-C}^2$ = $C_0$ (see
Fig~1b). (Note that in the non-catalysed case another degree of
freedom is also removed ($r_{A-D}^2$ - $r_{D-C}^2$ = $C_1$, see
Fig~1a) and an energy surface as a function of $C_0$, $C_1$ found.

In this mechanism the SW barrier to BF formation from the nearby
$C_{2v}$ isomer is decreased from 4.7~eV to 1.1~eV. Only one bond is
broken and one formed, as opposed to non-catalysed SW where there are
two such bond pairs. Another perspective is that the ad-atom has
already dilated the pentagon, part-transforming it to the hexagon
required in the product.

In the starting structure the adatom B sits proud of the BF cage above
a pentagon-hexagon bond. It rotates about this bond, flattening into
the BF cage and beginning to form a bond with host atom A. While it
does this, the bond from A to host atom C breaks and C lifts proud to
become the new adatom.

Overall the reaction is exothermic by 1.2~eV compared with 1.5~eV
without the adatom. The increased stability of the adatom on the
C$_{2v}$ paired pentagon system and the metastability of the final
C$_{61}$ state accounts for the difference. This metastable state of
the adatom, where it sits above a bond shared between a pentagon and a
hexagon (the 'homo' isomer of C$_{61}$), is 0.26~eV above its ground
state above a hexagon-hexagon bond (the 'methano' isomer of C$_{61}$).
This metastability is one reason why the mechanism was previously
overlooked.  Positive charging and high spin ({\it e.g.}
triplet) states cause only marginal increases in the activation barriers.

It is known that DFT(LSDA) calculations can underestimate reaction barriers
however this is typically of the order of 0.2~eV\cite{sadhukhan} and
thus will not qualitatively affect these results.  In addition any
error will systematically affect all calculated barriers including the
uncatalysed.

The mechanism, with its concomitant interstitialcy diffusion, is
entirely consistent with experiments in which a beam of $^{13}$C ions 
impinging on $^{12}$C C$_{60}$ molecules led to an output atom beam of
isotopic ratio of 1:60 $^{13}$C:$^{12}$C atoms\cite{canadians}.  

The observation that the adatom sits preferentially on defective
regions still applies \cite{ECSpaper}. We note that similar mechanisms
with vacant sites, rather than adatoms, can apply, but vacancies are
unlikely to migrate as easily.

Addressing the generalisation to other carbon structures, we
considered the annihilation of incipient glide dipole loops in
graphite and the motion of prismatic dislocations. In the polycyclic
aromatic hydrocarbon C$_{62}$H$_{20}$ we converted four hexagons into
two pentagon/heptagon pairs (a glide dislocation dipole, see Fig
2b). The dislocations are characterised by line directions $\pm
\langle$0001$\rangle$ and Burgers vector
$\langle2\bar{1}\bar{1}0\rangle$.  The SW barrier to annihilation was
found to be 5.6~eV and 0.7~eV, respectively for the uncatalysed and
adatom catalysed reactions (see Fig 3). The incipient dipole loop had
formation energies of 3.4~eV and 1.6~eV respectively. What is striking
is that the adatom catalysed barrier to remove the defective region is
very low and that there is a pronounced pull on the adatom to this
region.

Experience with the original autocatalysis suggests that the
activation barriers broadly scale with local curvature \cite{ECSpaper}
and thus these transformations within the shells of carbon onions (and
nanotubes) are expected to lie in the range of barriers calculated for
BF and graphite. Thus, since carbon onions are produced under
irradiation conditions, the mechanism explains the facility with which
spherical shape can be assumed, through changes in ring statistics
\cite{heggie+terrones_98}.

Interestingly, dimer addition to a nanotube followed by high uniaxial
tension has been proposed theoretically\cite{orlikowsky} as a means of
facilitating dislocation glide.  In the region through which glide has
occurred, the bonding topology, and hence semiconducting properties,
of the nanotube are altered giving rise to metal:semiconductor
junctions (and hence the possibility of a single nanotube device).
Our mechanism lends credence to this possibility - although we note that 
these authors did not find the reaction path we propose here.

Edge dislocations with ${\bf l} = \langle 0001 \rangle$ and ${\bf b} =
\langle 2\bar{1}\bar{1}0 \rangle$ have glide planes defined by 
${\bf b}$ x ${\bf l}$ which may lie between atoms closely spaced in
this direction or between atoms separated by a full bond length.
These two parallel planes we label the 'glide' and 'shuffle' planes,
respectively, because of their broad similarity with these planes in
the diamond structure in $\langle 1 1 0 \rangle$
projection\cite{LehtoHeggie}.

In the case of a graphene sheet, creation of a glide dislocation may
be accompanied by complete reconstruction of its core (Fig 2a),
(yielding a pentagon/heptagon pair) and the shuffle dislocation has an
extra atom inserted between this pair (Fig 2b).  This atom is bound by
2.29~eV to the core and is not differentiable from our adatom
models. Moving the glide dislocation is extremely difficult due to its
strong reconstruction (barrier 7.64 eV, applying a small correction
for edge effects).
However the shuffle dislocation (glide+adatom) moves by the 'new'
mechanism and has a much lower barrier of 2.22 eV.
Interestingly, this migration barrier is comparable with that in a
covalent semiconductor such as silicon, which is ductile above
approximately 700$^o$C : the temperature at which carbon onions
achieve near spherical shape under irradiation \cite{banhart_99}.

In conclusion, we have demonstrated a general mechanism by which
carbon structures evolve at moderate homologous temperatures in the
presence of a flux of adatoms.  The new reaction pathway presented
here gives activation barriers within the range 0.7 to 2.3 eV.  This
is around a factor of four smaller than equivalent uncatalysed
barriers and leads to qualitatively different behaviour.  In the case
of dislocations this may be lowered further by applied stress.

We acknowledge computing support from the Sussex High Performance
Computing Initiative and e6 consortium 'Ab Initio simulation of
Covalent Materials'. We also acknowledge fruitful discussion with
Prof. R. Jones.


\begin{figure}
\begin{center}
\ \psfig{file=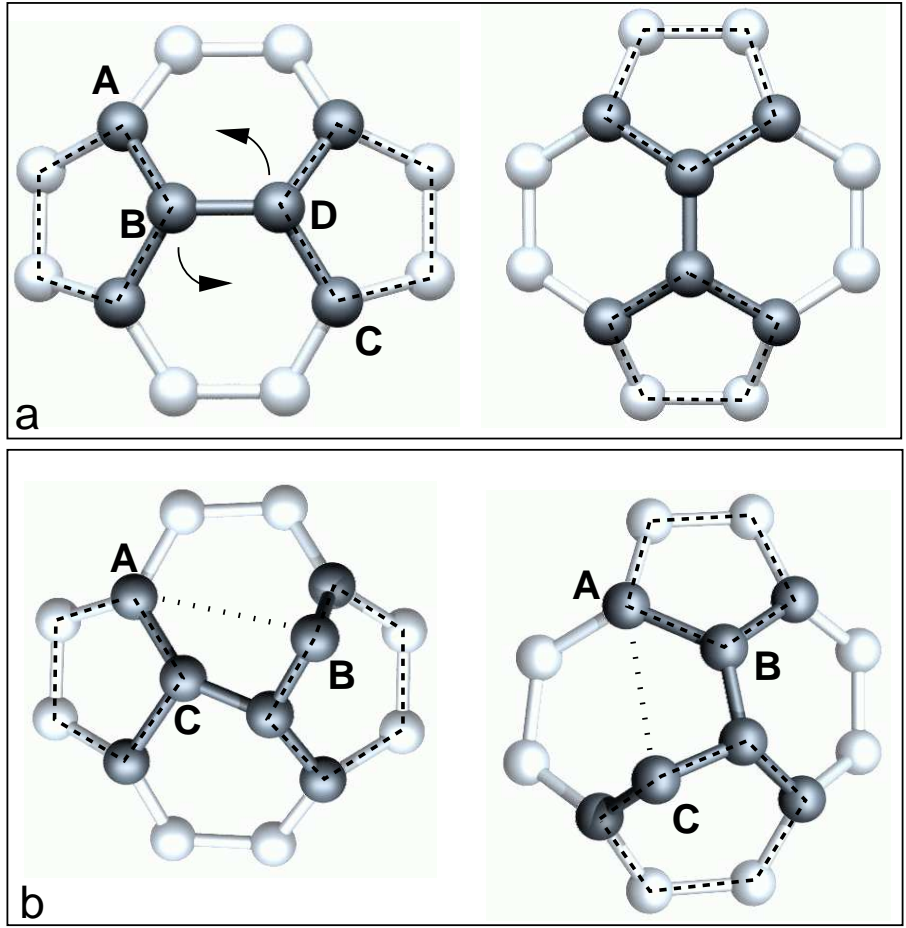,width=12cm}
\end{center}
\caption{Reaction pathway for Stone-Wales transformation (a) uncatalyed 
and (b) adatom catalysed. Pentagons are marked with dotted lines.}
\label{fig:c61}
\end{figure}

\begin{figure}
\begin{center}
\ \psfig{file=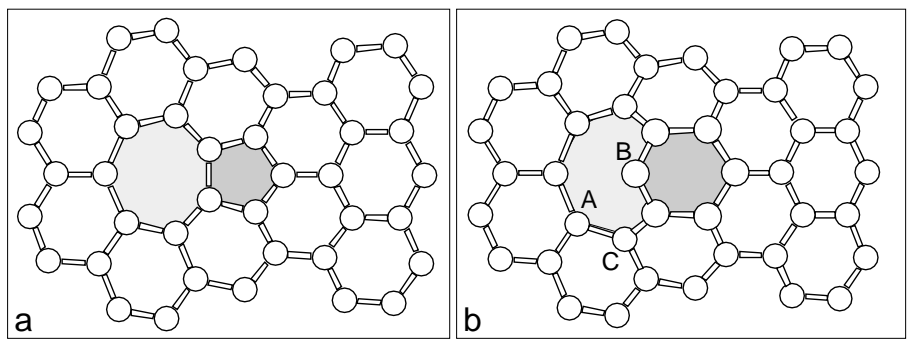,width=12cm}
\end{center}
\caption{Basal plane dislocations in graphite, (a) glide and 
(b) shuffle dislocation, equivilent to the glide with an adatom
bound at the core. A,B,C show atoms involved in motion as in Figure 1.}
\label{fig:disl_I}
\end{figure}

\begin{figure}
\begin{center}
\ \psfig{file=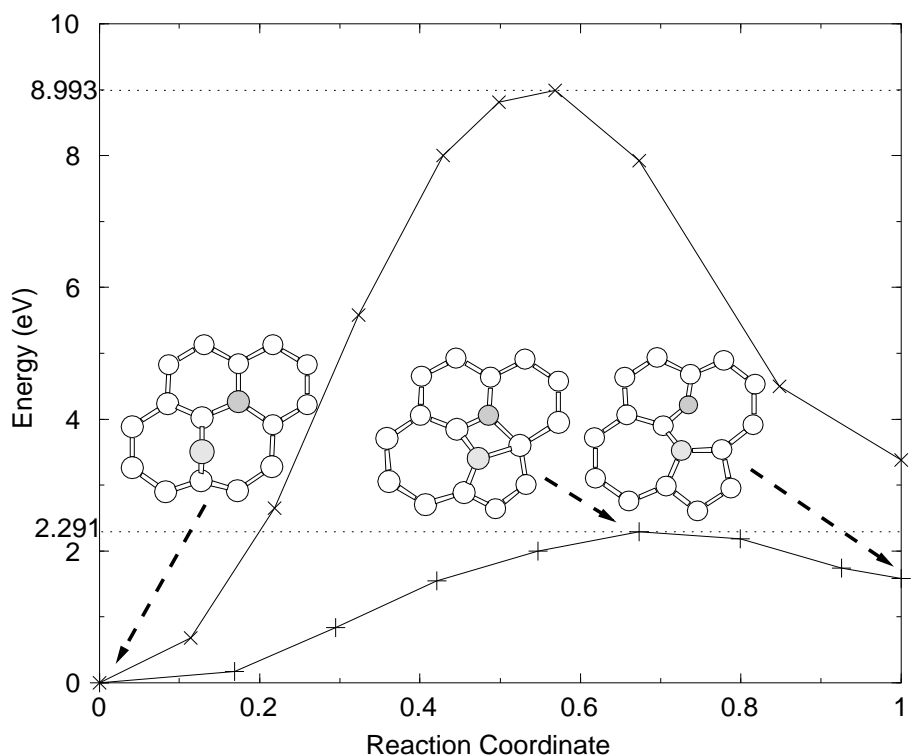,width=12cm}
\end{center}
\caption{Activation barrier (eV) for forming a glide dislocation dipole in graphene, with and without adatom catalysis.}
\label{fig:dip_bar}
\end{figure}



\end{document}